# Время работы системы с двумя различными ненадежными обслуживающими приборами.

The system operating time with two different unreliable servicing devices.

Головастова Э.А.

**Аннотация**

В данной статье изложено решение задачи теории надежности. Рассматривается система с двумя различными обслуживающими приборами, предполагается, что только один прибор находится в рабочем состоянии. Второй включается в работу после поломки первого, который немедленно отправляется на восстановление. Считаем известными функции распределения времени работы и восстановления обоих приборов. Цель работы - получение преобразования Лапласа для случайной величины, задающей время работы системы.

## 1. Задача надежности с двумя приборами.

Рассматривается система с двумя различными обслуживающими приборами в предположении, что обслуживание обеспечивается одним прибором. Оставшийся же включается в работу после поломки первого, который в это время отправляется на восстановление. Если сломавшийся прибор не успевает восстановиться за время работы другого, то считается, что система завершает свое функционирование, в противном, работа системы продолжается: восстановившийся прибор приступает к обслуживанию сразу же после поломки предшествующего и т.д. Заданными считаются функции распределения времени работы восстановления обоих приборов. Цель - получение преобразования Лапласа для случайной величины, задающей время работы системы.

Итак, есть 2 прибора, работающие последовательно, т.е. после поломки первого второй немедленно приступает к работе, первый же в это время восстанавливается. Если первый не успел восстановиться до поломки второго, то система прекращает работу, в противном цикл повторяется.

Приборы различны, времена работы и восстановления – независимые случайные величины.

Даны их функции распределения: работа: $F_1(x)$ $F_2(x)$, ремонт: $G_1(x)$ $G_2(x)$.

Найдем преобразование Лапласа времени работы системы. Пусть:

$\tau$ -время работы системы

$\varsigma$ - время работы прибора

$\eta$ - время ремонта прибора

$\varsigma_j^i$ - тут i – номер прибора ( 1-ый или 2-ой )

j – номер рабочего периода для i-ого прибора

(аналогично для ремонта)

## 1.1 Выражение для времени работы системы.

*Рассмотрим моменты отказа системы, когда за время работы 2-ого не успевает восстановиться 1-ый.*

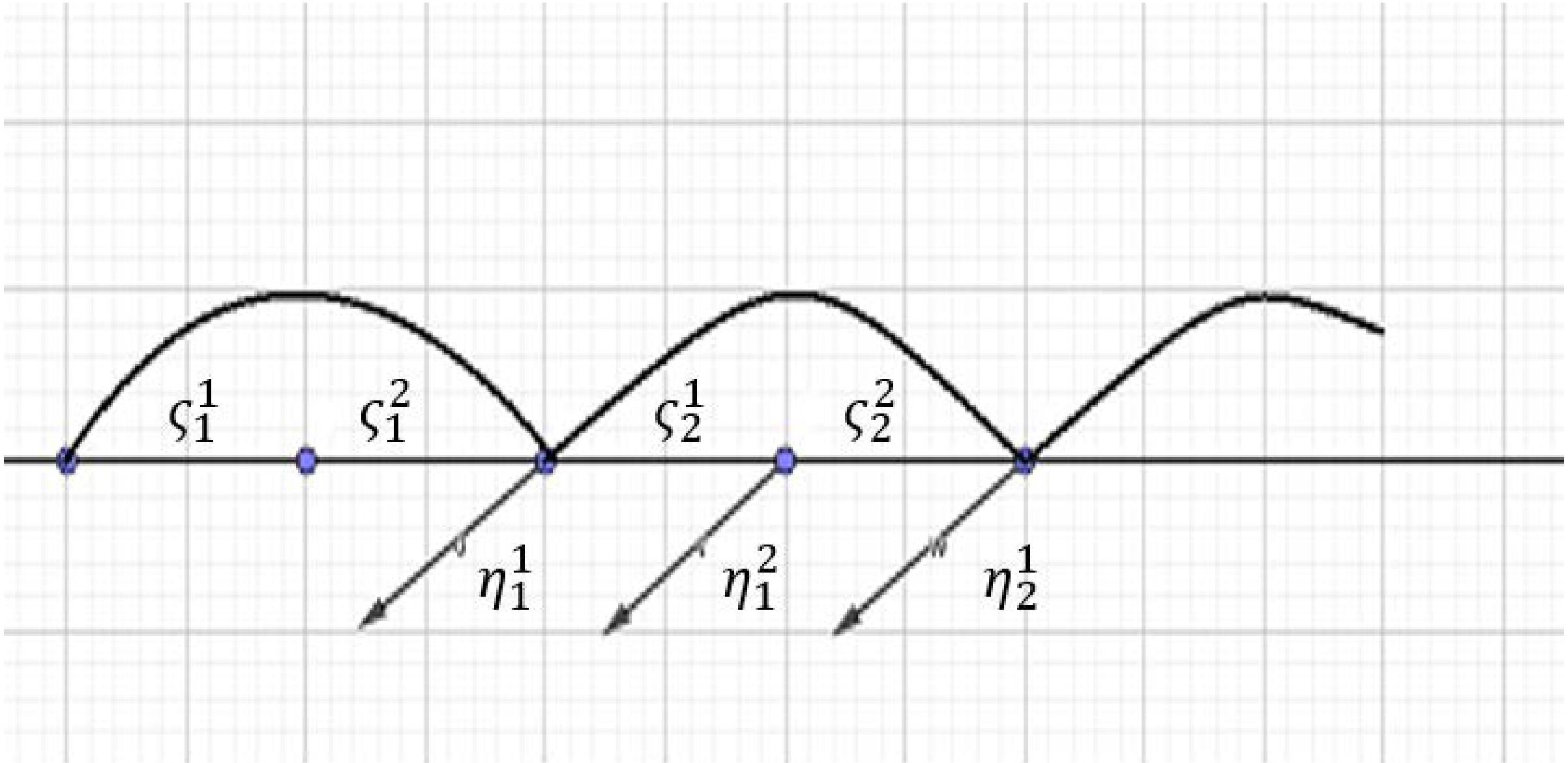

Рис.1.

Время работы системы:

$(\varsigma_1^1+\varsigma_1^2)$ $\qquad \eta_1^1>\varsigma_1^2$

$(\varsigma_1^1+\varsigma_1^2)+(\varsigma_2^1+\varsigma_2^2)$ $\qquad \eta_1^1<\varsigma_1^2;\eta_1^2<\varsigma_2^1;\eta_2^1>\varsigma_2^2$

…

$$(\varsigma_1^1+\varsigma_1^2)+(\varsigma_2^1+\varsigma_2^2)+\ldots+(\varsigma_k^1+\varsigma_k^2)$$

$\eta_1^1<\varsigma_1^2;$

$\eta_1^2<\varsigma_2^1;\eta_2^1<\varsigma_2^2;$

$\eta_2^2<\varsigma_3^1;\eta_3^1<\varsigma_3^2;$

$\eta_{k-1}^2<\varsigma_k^1;\eta_k^1>\varsigma_k^2;$

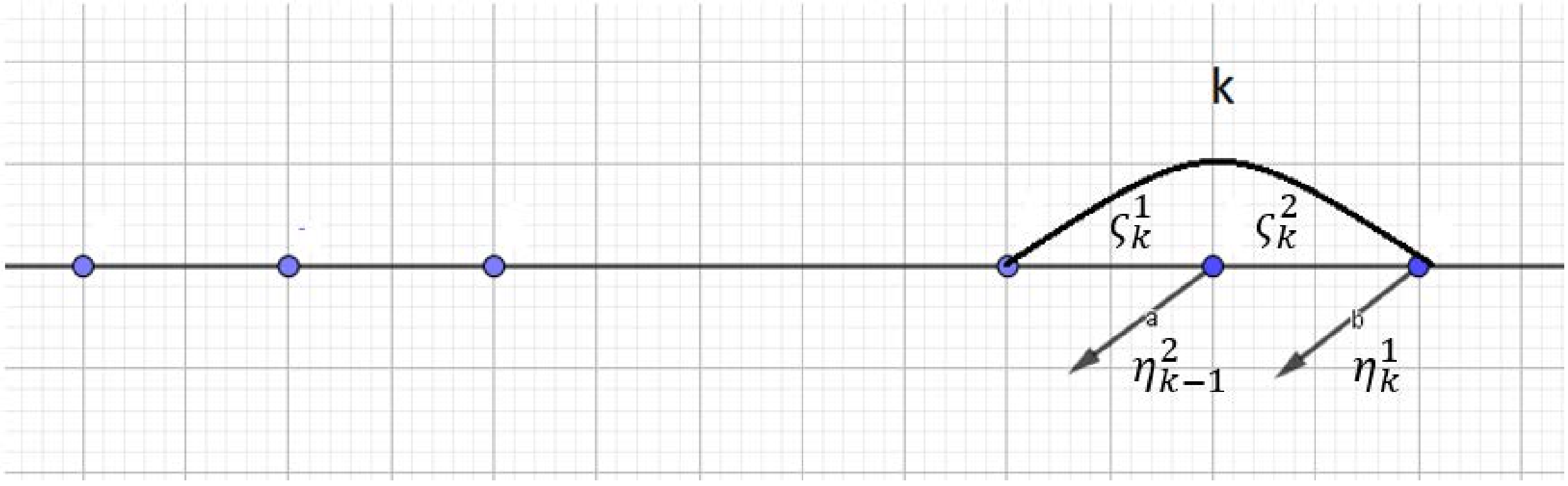

Рис.2.

*Рассмотрим моменты отказа системы, когда за время работы 1-ого не успевает восстановиться 2-ый.*

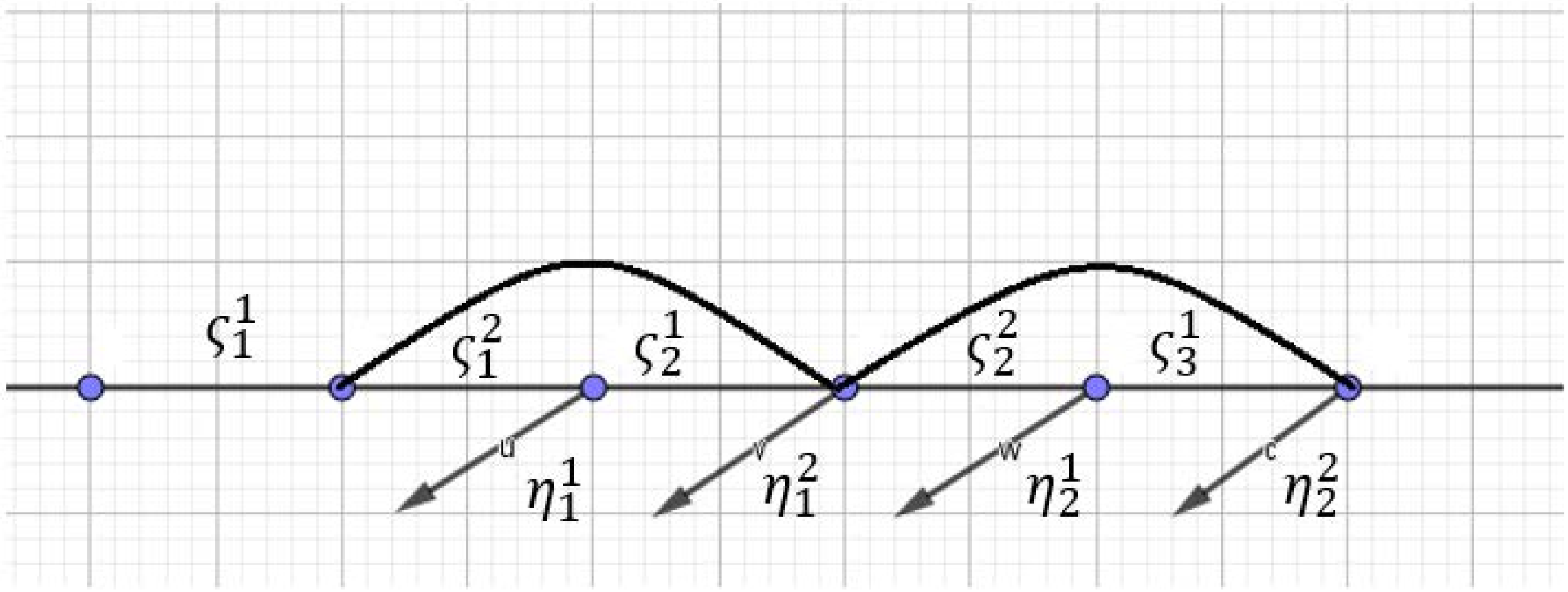

Рис.3.

Время работы системы:

$\varsigma_1^1+(\varsigma_1^2+\varsigma_2^1)$ $\qquad$ $\eta_1^1<\varsigma_1^2;\eta_1^2>\varsigma_2^1$

$\varsigma_1^1+(\varsigma_1^2+\varsigma_2^1)+(\varsigma_2^2+\varsigma_3^1)$ $\qquad$ $\eta_1^1<\varsigma_1^2;\eta_1^2<\varsigma_2^1;$ $\eta_2^1<\varsigma_2^2;\eta_2^2>\varsigma_3^1$

…

$$\varsigma_1^1 + (\varsigma_1^2 + \varsigma_2^1) + \ldots + (\varsigma_k^2 + \varsigma_{k+1}^1)$$

$$\eta_1^1 < \varsigma_1^2; \eta_1^2 < \varsigma_2^1;$$
$$\eta_2^1 < \varsigma_2^2; \eta_2^2 < \varsigma_3^1;$$
$$\eta_k^1 < \varsigma_k^2; \eta_k^2 > \varsigma_{k+1}^1$$

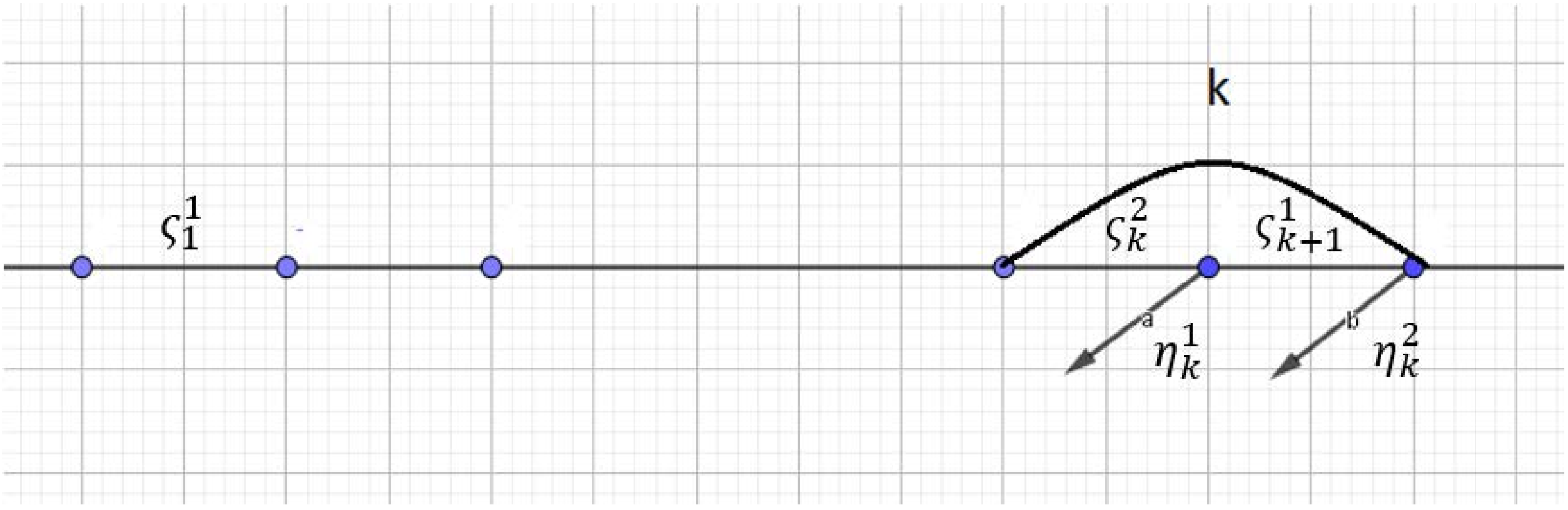

Рис. 4.

Тогда преобразование Лапласа времени работы системы:

$$Ee^{-s\tau} = E\{e^{-s(\varsigma_1^1+\varsigma_1^2)} \cdot I(\eta_1^1 > \varsigma_1^2) +$$

$$+\sum_{k=2}^{\infty} e^{-s[\sum_{i=1}^{k}(\varsigma_i^1+\varsigma_i^2)]} \cdot I(\eta_k^1 > \varsigma_k^2; \eta_j^1 < \varsigma_j^2; \eta_j^2 < \varsigma_{j+1}^1; j = 1 \ldots k-1) +$$

$$+\sum_{k=1}^{\infty} e^{-s\varsigma_1^1} \cdot e^{-s[\sum_{i=1}^{k}(\varsigma_i^2+\varsigma_{i+1}^1)]} \cdot$$

$$\cdot I(\eta_k^2 > \varsigma_{k+1}^1; \eta_k^1 < \varsigma_k^2; \eta_j^1 < \varsigma_j^2; \eta_j^2 < \varsigma_{j+1}^1; j = 1 \ldots k-1)\} = 1) + 2) + 3)$$

Пусть

$$f^1 = Ee^{-s\varsigma^1}$$
$$f^2 = Ee^{-s\varsigma^2}$$

$$\phi^1 = Ee^{-s\varsigma^1} \cdot I(\varsigma^1 > \eta^2) = \int_0^{\infty} e^{-sx} G_2(x) dF_1(x)$$

$$\phi^2 = Ee^{-s\varsigma^2} \cdot I(\varsigma^2 > \eta^1) = \int_0^{\infty} e^{-sx} G_1(x) dF_2(x)$$

Тогда

$$1) = f^1 \cdot (f^2 - \phi^2)$$

$$2) = E\{\sum_{k=2}^{\infty} e^{-s\varsigma_1^1} \cdot \prod_{j=1}^{k} e^{-s\varsigma_j^1} \cdot I(\varsigma_j^1 > \eta_{j-1}^2) \cdot$$

$$\cdot \prod_{j=1}^{k-1} e^{-s\varsigma_j^2} \cdot I(\varsigma_j^2 > \eta_j^1) \cdot$$

$$\cdot e^{-s\varsigma_k^2} \cdot I(\eta_k^1 > \varsigma_k^2)\} =$$

$$= f^1 \cdot (f^2 - \phi^2) \cdot \sum_{k=2}^{\infty} (\phi^1)^{k-1} \cdot (\phi^2)^{k-1}$$

$$3) = f^1 \cdot E\{\sum_{k=1}^{\infty} e^{-s\varsigma_k^2} \cdot I(\eta_k^1 < \varsigma_k^2) \cdot e^{-s\varsigma_{k+1}^1} \cdot I(\eta_k^2 > \varsigma_{k+1}^1) \cdot$$

$$\cdot \prod_{j=1}^{k-1} e^{-s\varsigma_j^2} \cdot I(\varsigma_j^2 > \eta_j^1) \cdot$$

$$\cdot \prod_{j=2}^{k} e^{-s\varsigma_j^1} \cdot I(\varsigma_j^1 > \eta_j^2)\} =$$

$$= f^1 \cdot \phi^2 \cdot (f^1 - \phi^1) \cdot \sum_{k=1}^{\infty} (\phi^1)^{k-1} \cdot (\phi^2)^{k-1}$$

Тогда

$$Ee^{-s\tau} = f^1 \cdot (f^2 - \phi^2) \cdot \sum_{k=0}^{\infty} (\phi^1 \phi^2)^k +$$

$$f^1 \cdot \phi^2 \cdot (f^1 - \phi^1) \cdot \sum_{k=0}^{\infty} (\phi^1 \phi^2)^k =$$

$$= \frac{f^1}{1 - \phi^1 \phi^2} \cdot (f^2 - \phi^2 + \phi^2 \cdot (f^1 - \phi^1))$$

## 1.2 Иной способ.

Данную задачу можно решить другим способом. Пусть:

$\tau_{12}$ - время до поломки системы на момент заступления 1-ого прибора на работу и ухода 2-ого на ремонт.

$\tau_{21}$ - наоборот.

$$g_{12} = Ee^{-s\tau_{12}}$$

$g_{21}$ - аналогично.

Если сначала работает 1-ый прибор, то общее время работы системы $\tau$ :

$$Ee^{-s\tau} = f^1 \cdot g_{21} = Ee^{-s(\varsigma^1+\tau_{21})}$$

$$\tau_{12} = \varsigma^1 \cdot I(\varsigma^1 \le \eta^2) + (\varsigma^1 + \tau_{21}) \cdot I(\varsigma^1 > \eta^2)$$
$$\tau_{21} = \varsigma^2 \cdot I(\varsigma^2 \le \eta^1) + (\varsigma^2 + \tau_{12}) \cdot I(\varsigma^2 > \eta^1)$$

Заметим, что $\tau_{12}$ и $\tau_{21}$ ни от чего не зависят в правой части равенства.

$$Ee^{-s\tau_{12}} = E\{e^{-s\varsigma^1} \cdot I(\varsigma^1 \le \eta^2)\} + E\{e^{-s(\varsigma^1+\tau_{21})} \cdot I(\varsigma^1 > \eta^2)\}$$
$$Ee^{-s\tau_{21}} = E\{e^{-s\varsigma^2} \cdot I(\varsigma^2 \le \eta^1)\} + E\{e^{-s(\varsigma^2+\tau_{12})} \cdot I(\varsigma^2 > \eta^1)\}$$

Тогда во введенных обозначениях последние уравнения перепишутся:

$$g_{12} = (f^1 - \phi^1) + \phi^1 \cdot g_{21}$$
$$g_{21} = (f^2 - \phi^2) + \phi^2 \cdot g_{21}$$

И тогда:

$$g_{12} = \frac{f^1 - \phi^1 + \phi^1 (f^2 - \phi^2)}{1 - \phi^1 \phi^2}$$
$$g_{12} = \frac{f^2 - \phi^2 + \phi^2 (f^1 - \phi^1)}{1 - \phi^1 \phi^2}$$

## Список литературы: